\documentclass[11pt]{amsart}

\usepackage{url}
\usepackage{graphicx}

\title[On Creativity of Slime Mould]{On Creativity of Slime Mould} 

\author[Adamatzky]{Andy Adamatzky$^1$}
\author[Armstrong]{Rachel Armstrong$^2$}
\author[Jones]{Jeff Jones$^1$}
\author[Gunji]{Yukio-Pegio Gunji$^3$}

\address[Adamatzky]{University of the West of England, Bristol, United Kingdom}
\address[Armstrong]{University of Greenwich, London, United Kingdom}
\address[Jones]{University of the West of England, Bristol, United Kingdom}
\address[Gunji]{University of Kobe, Kobe, Japan}

\begin{document}

\maketitle

\centerline{$^1$ \footnotesize{University of the West of England, Bristol, United Kingdom}}
\centerline{$^2$ \footnotesize{University of Greenwich, London, United Kingdom}}
\centerline{$^3$ \footnotesize{University of Kobe, Kobe, Japan}}

\begin{abstract}
Slime mould \emph{Physarum polycephalum} is large single cell with intriguingly smart behaviour. The slime mould shows outstanding abilities to adapt its protoplasmic network to varying environmental conditions.  The slime mould can solve tasks of computational geometry, image processing, logics and arithmetics when data are represented by configurations of attractants and repellents.  We attempt to map  behavioural patterns of slime onto the cognitive control versus schizotypy spectrum phase space and thus interpret slime mould's activity in terms of creativity.

\vspace{0.5cm}


\end{abstract}

\section{Introduction}

This paper represents a multi-disciplinary approach to considering the 
way that non-human actants~\footnote{`Actant' is Bruno Latour's term for a source of action; an actant can be human or not, or, most likely, a combination of both. It implies no special motivation of human individual actors, nor of humans in general~\cite{bennett_2010,latour_1996}} make choices and potentially express 
`creativity' in their actions, in ways that are not accounted for by 
conventional Western schools of philosophical thought. The importance of 
this investigation is in its exploration of issues that may help 
articulate the production of novelty and the underlying by material 
systems that are dynamic yet do not possess a formal nervous system such 
as slime mould \emph{Physarum polycephalum}. Our choice of model enables an exploration and 
reflection of important issues that relate to the production of novelty 
and the kinds of `actants' that may constitute a creative `assemblage':  in the case of the slime mould 
this is environment, chemical artefacts and organism. 

As such, this paper provides an alternative method of 
exploring creativity to the traditional dualism that exists between the 
purposeless of `random' events that underpin novelty and evolution, or
  the `intelligent designer', or `vitalisitc' forces that direct 
emergence. This paper proposes that `creativity' can be observed in 
nature as the deliberate (but not rational) interactions between 
chemical processes (agents) and is the outcome of the restless condition 
that is innate to matter and which resides in the realm of quantum 
physics, which is yet to be fully explored.

\section{`Intelligence' of slime moulds and morphological `meaning'}

\emph{Physarum polycephalum} belongs to the species of order \emph{Physarales}, subclass \emph{Myxogastromycetidae}, class \emph{Myxomycetes}, division \emph{Myxostelida}. It is commonly known as a true, acellular or multi-headed slime mould. \emph{P. polycephalum} has a complex life cycle.  Plasmodium is a 'vegetative' phase, a single cell with a myriad of diploid nuclei. It is visible to the naked eye and looks like an amorphous yellowish mass with networks of protoplasmic tubes. The plasmodium behaves and moves as a giant amoeba. It feeds on bacteria, spores and other microbial creatures and micro-particles~\cite{stephenson_2000}. When foraging for its food the plasmodium propagates towards sources of food particles, surrounds them, secretes enzymes and digests the food. 

\begin{figure}[!tbp]
\centering
\includegraphics[width=0.8\textwidth]{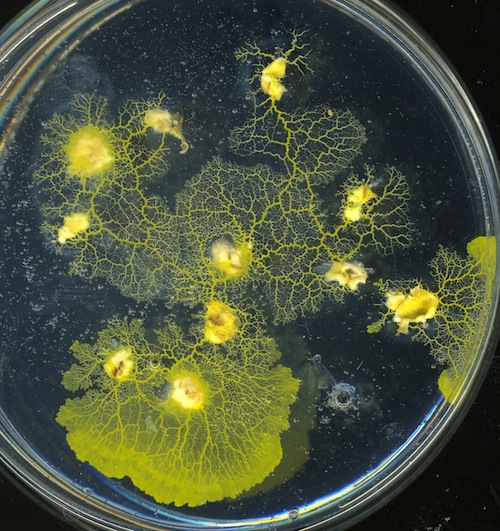}
\caption{Plasmodium of \emph{P. polycephalum} colonises oat flakes on a non-nutrient agar gel. }
\label{fig01}
\end{figure}

Typically, the plasmodium forms a network of protoplasmic tubes connecting the masses of protoplasm at the food sources. These networks have been shown to be efficient in terms of length and resilience~\cite{nakagaki_2001}. When several sources of nutrients are scattered in the plasmodium's range, the plasmodium forms a network of protoplasmic tubes connecting the masses of protoplasm at the food sources (Fig.~\ref{fig01}).

The plasmodium is a network of biochemical oscillators~\cite{matsumoto_1988}. Waves of excitation or contraction originate from several sources, e.g. induced by external stimuli and perturbations. The waves travel along the plasmodium and interact one with another in collisions. The oscillatory cytoplasm of the plasmodium is a spatially extended nonlinear excitable medium. Growing and feeding plasmodium exhibits characteristic rhythmic contractions with articulated sources. The contraction waves are associated with waves of electrical potential change. The waves observed in plasmodium~\cite{yamada_2007}are similar to the waves found in excitable chemical systems, such as Belousov-Zhabotinsky medium. The following wave phenomena were discovered experimentally~\cite{yamada_2007}: undisturbed propagation of contraction waves inside the cell body, collision and annihilation of contraction waves, splitting of the waves by inhomogeneity, and the formation of spiral waves of contraction. These phenomena closely match the dynamics of pattern propagation in excitable reaction-diffusion chemical systems~\cite{adamatzky_2005,adamatzky_physarummachines}. 

Plasmodium's foraging behaviour can be interpreted as a computation~\cite{adamatzky_physarummachines} as follows.  Data are represented by spatial configurations of attractants and repellents.  Results are represented by the structure of the protoplasmic  network.  
Plasmodium can solve computational problems with natural parallelism, e.g. related to shortest 
path~\cite{nakagaki_2001} and hierarchies of planar proximity graphs~\cite{adamatzky_ppl_2008}, computation of plane tessellations~\cite{adamatzky_physarummachines,shirakawa, shirakawa_2011}, execution of logical computing schemes~\cite{tsuda_2004, adamatzky_parco}, planar shapes and concave hulls~\cite{adamatzky_parco}, 
and natural implementation of spatial logic and process algebra~\cite{schumann_adamatzky_2009}. 

The computational behaviour of the plasmodium is notable for two features. Firstly is the fact that the organism is not explicitly trying to solve computational problems. It does not comprehend or care about such problems, it is simply just trying to survive. Therefore the previously noted computational feats must be represented in a manner that is a good computational `fit' between the natural interactions of the organism and its environment. The interpretation of plasmodium behaviour as computation is externally applied, although, with the wide body of recent evidence as noted in the previous examples, cannot be doubted. The second feature of interest is that the mechanisms by which the plasmodium carries out its computation are qualitatively different from those observed either in the symbolic logic of classical computing or in the neural based systems in higher-order animals. The range of different computational feats performed by this single cell is remarkable given its relative material simplicity. 

The wide range of computation, and novel mechanisms employed, render \emph{Physarum} plasmodium as a suitably minimal candidate to examine concepts of logic, problem solving, association, memory, decision-making and combinatorial optimisation. Many of these concepts have hitherto been restricted to humans. The fact that the plasmodium is constructed out of such simple components suggests that these higher-order concepts may arise from similarly simple and local interactions and need not reside within any `special' structure or component. The aim of such constructive examination is not intended to inflate the abilities of the humble plasmodium, nor to reduce disparage the importance of these concepts. Rather it is to try and discover the possible mechanical `building-blocks' and environmental interactions which may generate these high-level phenomena.

The plasmodium expresses itself via morphology of its body and its protoplasmic network. Can the topology of protoplasmic tubes be interpreted in terms of creativity? Yes, via `morphological meaning'.

Essentially, while we traditionally deal with `creativity' as being something internally generated (like logic) to consider an environmental role requires the attribution of `meaning' to context, which can be explored by the agent, or organism (in this case).  But how can we ÔdemonstrateÕ the meaning within systems?

Currently, empirical measurements are employed to address the qualities of matter that Aristotle was so fond 
of~\cite{rosen_1996}.  In this context, the term `meaning' could be used to describe the contingent entanglement of qualities and forces that compel `actants' (or organisms) `to keep on existing'~\cite{latour_2012}.
The philosophical understanding of `meaning' is traditionally equated with human cognition.  We are not using this term in its anthropocentric context but in reference to Bruno Latour's notion of non-human actants that exert an autonomous, innate force as a consequence of their being.

In this context `meaning' (the contingent entanglement of qualities and forces) can be understood as a transformer of information flow in the system, which converts `risk' into action. Rather than using human semiotics, these systems exchange meaning through other modes, which may be found in chemistry or systems that do not possess an organised nervous system. For example, bacteria use chemical signaling to interact with each other, which has been described as a kind of `language' that uses `words'~\cite{schauder_2001,bassler_2012}, can be listened to \cite{cataldi_2012},
and extends between the bacterial realm and can influence non-bacterial actants~\cite{pacheco_2009}.

The grammar of a chemical language may be forged through the parallel processing exchanges of `unconventional computation' that underpins the flow of  information and `meaning' between actants to generate `risk-taking' actions. The notion of `risk taking' is in keeping with Ilya Prigogine's observation that matter exists in a probabilistic, not deterministic state~\cite{prigogine_1997} and as a consequence of a set of contingencies (meaning) can therefore exert force, or produce effects, without needing to be directed by a `conscious' entity.  In this sense `meaning' is the `transformer' through which actants employ risk-taking, editing systems to respond to their context. Some of these interactions are sophisticated and operate through behavioural modifications, while others are purely physical/chemical and rely on the breaking or forming of molecular bonds.  However, increasingly the sole agentism of this theory is being challenged, specifically with respect to how DNA itself is regulated~\cite{keller_2000} and 
the `excruciating intractability'~\cite{gould_2002}  of the cell matrix from its environment~\cite{keller_1983}.

Eva Jablonka and Marion J. Lamb (2006) characterise alternative mechanisms to the `random mutation' of DNA proposed by the Neo-Darwinists that also govern evolutionary mechanisms Ð namely: genetic, epigenetic (developmental), behavioural, and symbolic (linguistic).
Significantly, they propose that developmental, behavioural, and
 linguistic attributes are able to impinge on evolution through a process of assimilation, although Neo-Darwinists insist that all four dimensions are ultimately the consequence of genetic strategies~\cite{benton_2005}.  Yet, Jablonka and Lamb anticipate this criticism and make their argument using case studies to demonstrate how their different mechanisms affect the final phenotype of an organism, which may persist beyond a generation and how this contrasts with Neo-Darwinian genetics. 
 
  Meaning permeates David Bohm's Implicate Order~\cite{bohm_1980}  being present in bodies and environments. It is locally organized by actants operating through assemblages. Jane Bennett notes that actants that act weakly can amplify their force and effect through assemblages. ``An actant never really acts alone\ldots [it] depends on the collaboration, cooperation, or interactive interference of many bodies and forces''. ``Bodies enhance their power in or as a heterogeneous assemblage\ldots  Assemblages are ad hoc groupings of diverse elements, of vibrant materials of all sorts.''
  \cite{bennett_2010} These are not limited to humankind's semiotics, nor is it driven by any single agency.

\centerline{* \hspace{0.2cm} * \hspace{0.2cm} * \hspace{0.2cm} }

In this paper we examine the innate material behaviour of the \emph{Physarum} plasmodium in its interaction with the stimuli presented by its environment, relating these interactions to recent explorations which attempt to broadly encompass and tentatively categorise the nebulous concept of creativity. Can this simple organism provide clues by either mechanism or metaphor to the emergence of creativity?

\section{On creativity}

There is a myriad of ways to define and explore creativity. They include creativity as mesoscopic design proposed by Licata and Minati (2010);
 Kowaliw-Dorin-Korb creativity based on probability of pattern emergence~\cite{dorin_2009, kowaliw_2009}; notions of Stuart Kauffman's `radical creativity' of the universe~\cite{kauffman_2008}; Prigogine's ideas about the creativity of `time's arrow'~\cite{prigogine_1997}. There are also more cultural evaluations by Deleuze and Guattari (2004),
 Manuel De Landa (2011)
and Jane Bennett (2010) 
--- who are all arguing for a `vibrant materiality' in other words, matter with agency --- which sits outside of the traditional Western Cartesian philosophical perspective (which requires human agency to observe or sense it). Some might argue that the `creativity' of matter actually sits much better with gnosticism, more in keeping with Vernadsky's perspective of, the biogeosphere~\cite{vernadsky_1998}.  

Creativity may not correlate with intelligence, in humans correlation between IQ and creativity vanishes 
after IQ=120~\cite{guilford_1956, sternberg_1998, kuszewski_2009}. As Kuszewski (2009)
stated, referring to~\cite{runco_2006}, a high degree of intelligence can limit a flexibility of divergent thinking and thus diminish creativity. 
 
We understand that there is a distinction between creativity, which is loaded with subjective meaning, and the slime mould's behaviour, which observes performance objectively.  In layman's terms, to call something `creative' there needs to be a condition of culture, or a habituated context, against which any `innovation' can be judged by peers or through an objective assessment of performance, e.g. time to complete a task. Despite the challenges in interpretations, we are trying to offer some kind of working model which will be if not a solution, then a great starting points towards exploration of creativity in apparently primitive living substrates. 

Csikszentmihalyi (2006)
has shown that great creativity  ``the kind that changes some aspects of the culture, is never only in the mind of a person''. He postulated that creativity can be observed only in the interrelations of a system made up of three main parts: domain, field and person. The domain refers to the cultural system, which consists of a system of symbolic rules and procedures. The field includes all the gatekeepers of a given domain. The field determines what products are regarded as creative. The person is the actor in the system that actually uses the symbols in a giving domain and which ideas and products are chosen as innovative. 


We interpret domain as slime mould's foraging behaviour in terms of its creative involvement with morphological meaning in its surroundings. Where the `field' can be understood as the plasmodium's physical environment and person as the \emph{Physarum's} behavioural patterns. The domain can also be interpreted in terms of conceptual space as choices made between a set of artefacts~\cite{ritchie_2006}, using a chemical `language'. For \emph{Physarum}, this constitutes a configuration of sources of attractants and repellents and a continuous - and dynamically changing - field of chemical and physical gradients.

From a psychological and neurophysiological perspective there is a great similarity between creativity and 
psychoticism~\cite{eysenck_1992, eysenck_1993, abraham_2005, koh_2006, shermer_2011}. The similarities include over-inclusive cognitive style, conceptual expansion, associative thinking, lateral thinking dominates vertical (goal-oriented) thinking. In contrast to creativity, however,  psychoticism shows diminished practicality\cite{abraham_2005,koh_2006}. There are physiological evidences, see review in~\cite{kuszewski_2009}, that creative individuals show activities in both hemispheres and increased inter-hemispheric transfer.  As Salvador Dali employed the `paranoid critical method' in producing his work, deliberately entreating the subconscious mind to engage creatively with the world. Andre Breton and the surrealists used many forms of creative practice to evade rational `control' in their work using for example, `automatic drawing' techniques, game play (the Exquisite Corpse parlour game where a drawing is assembled by multiple players unaware of the contributions by others) and `frottage' a technique employed by Max Ernst which involved rubbing sheets of paper to pick up subtle, invisible traces in the environment which of course, were not directed by the rational mind.

In her refreshingly inspiring paper  Kuszewski (2009)
provides the following indicators of creativity. They are (1) divergent thinking and lack of lateral inhibition, (2) the ability to make remote associations between ideas and concepts,  (3) the ability to switch back and forth between conventional and unconventional ideations (flexibility in thinking), (5) to generate novel ideas appropriate for actualities, (6) willingness to take risks, (7) functional non-conformity.  Let us interpret these criteria in the framework of the slime mould behaviour.

\section{Divergent thinking}

\begin{figure}[!tbp]
\centering
\includegraphics[width=0.9\textwidth]{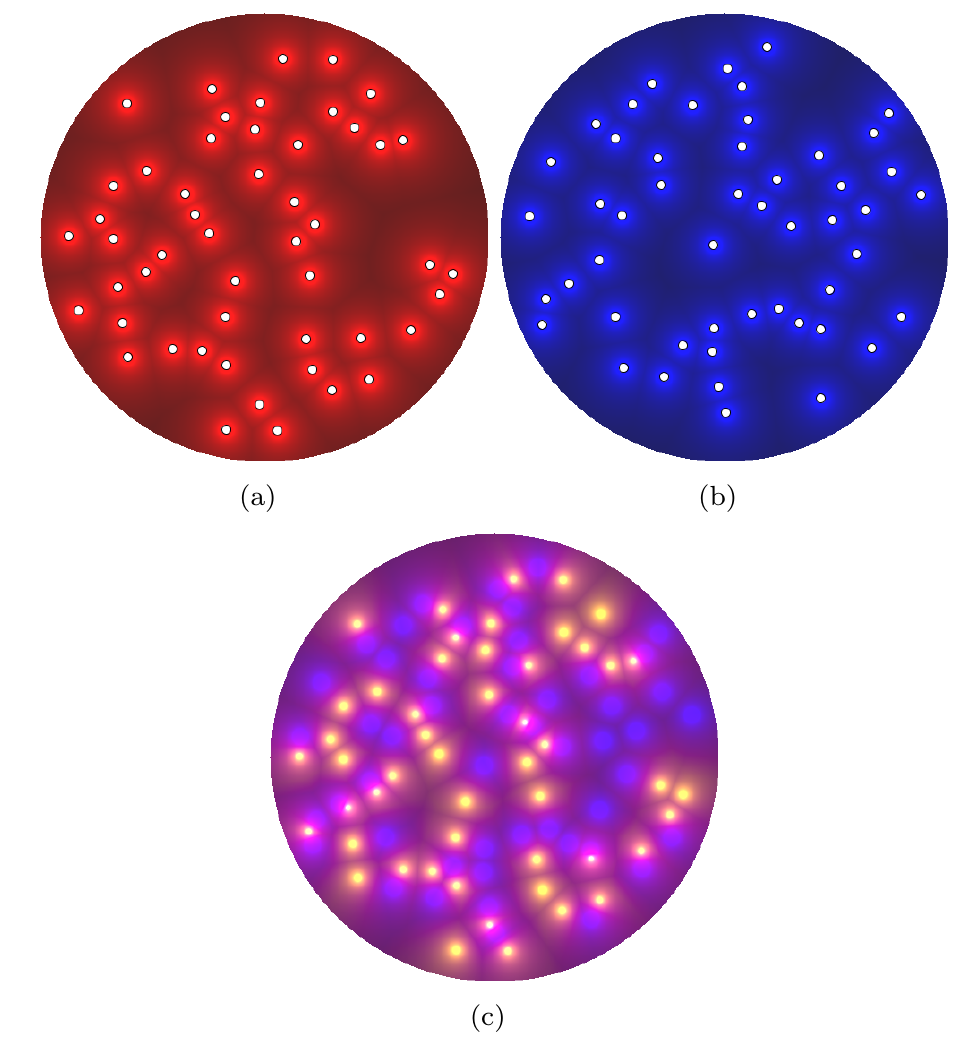}
\caption{How \emph{Physarum} `sees' the world: (a) attractant field, (b) repellent gradient field, (c) \emph{Physarum} spatial vision of its world as operations between attracting and repelling stimuli. }

\label{fig02}
\end{figure}

\emph{Physarum} lives in gradient fields of attractants and repellents (Fig.~\ref{fig02}).  Its thinking (or perception) is tactile, either physically or via chemo-receptors, and spatialised. When growing and colonising a substrate  \emph{Physarum} maximises its presence in domains with higher concentration of attractants and lower concentration of repellents. The divergent thinking of \emph{Physarum} is expressed in its simultaneous reaction to several sources of attractants and repellents, and parallel implementation of sensorial fusion

\begin{figure}[!tbp]
\centering
\includegraphics[scale=1]{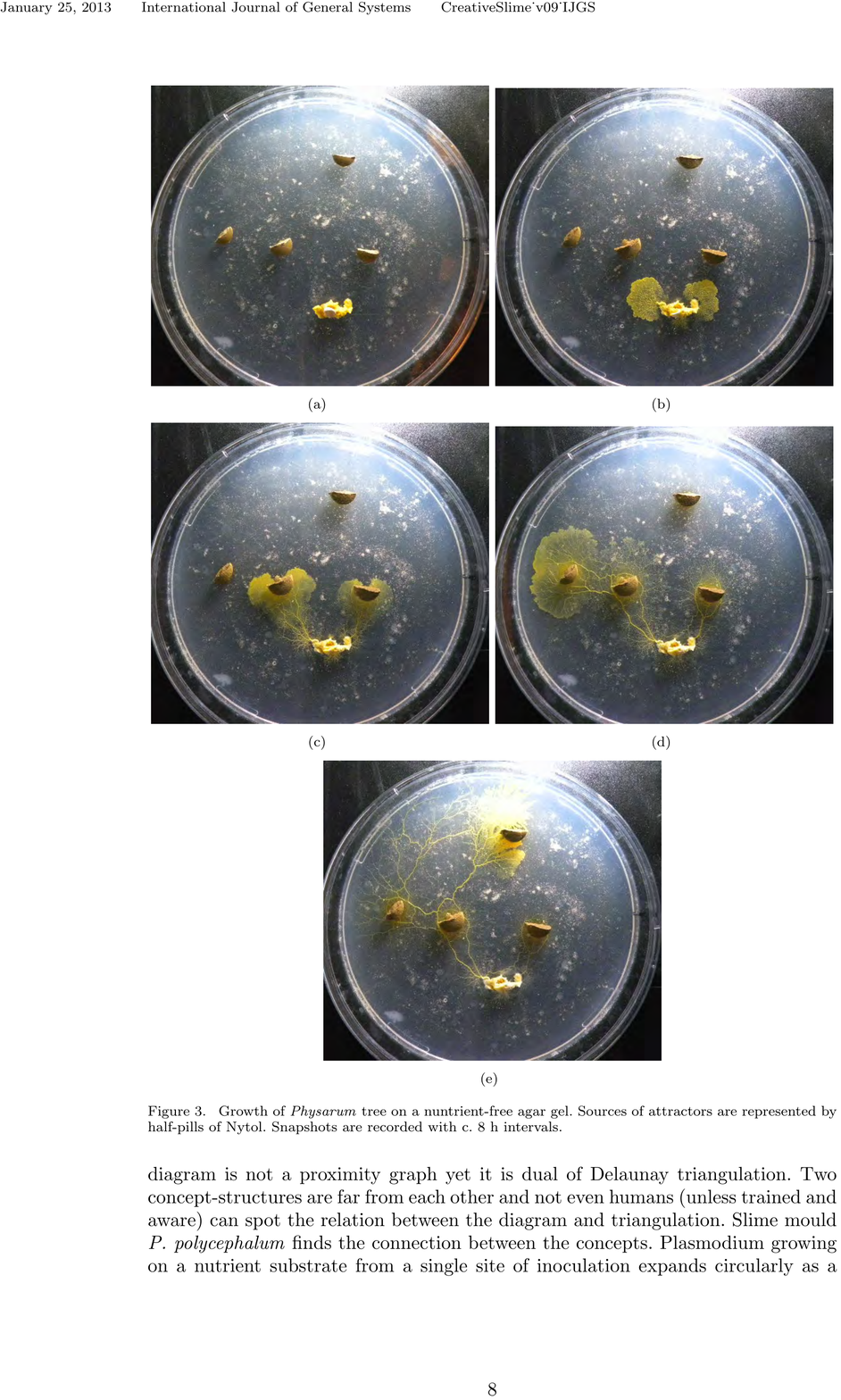}
\caption{Growth of \emph{Physarum} tree on a nuntrient-free agar gel. Sources of attractors are represented by 
half-pills of Nytol. Snapshots are recorded with c. 8~h intervals.}
\label{tree}
\end{figure}

A growing substrate is \emph{Physarum's} concept space. Sources of attractants are concepts. As illustrated in Fig.~\ref{tree} \emph{Physarum} explores several concepts at a time branching from one concept to another. At the beginning of experiments
\emph{Physarum} is inoculated in southern part of a Petri dish (Fig.~\ref{tree}a). At first it undergoes binary branching and propagates towards two most closest to it pills at once  (Fig.~\ref{tree}b). Then \emph{Physarum's} activity zones move towards westmost pill  (Fig.~\ref{tree}b) and then towards northmost pill (Fig.~\ref{tree}c).
  
Activity in both hemispheres and exchange of activities between hemispheres are considered to be attributes of human creativity (Kuszewski, 2009). In \emph{Physarum} the hemispheres' activity  is represented by simultaneous oscillatory activity, with biochemical oscillator located to distant parts of \emph{Physarum} body and oscillating with different frequencies and amplitudes. Interaction between hemispheres is instantiated by waves of contractile activity with associated changes in electrical potential propagating along protoplasmic tubes.

\section{Remote associations between concepts} 
\label{remoteassociations}

\emph{Physarum}  explores space and forages for  nutrients (meaningful ideas) and non-nutrient (not useful ideas) domains of space. Its ability to make remote associations between ideas is expressed in its physical response to spatial information via decision making and temporalization of its tactile thinking.  The plasmodium, although initially exploring in all directions, soon builds up some momentum in one particular direction. This momentum lends a certain inertia to the foraging of the plasmodium and it may, for example, maintain its directional persistence for a short time even after a stimulus is removed. The `attention' paid by the plasmodium in a complex environment with numerous stimuli is partially dependent on distance and chemo-attractant concentration, but is subject to influence by environmental noise. The mechanism behind such switches in association (or attention) may be the lag caused by disparity between the fast transmission of electrical potential and the relatively slow bulk transfer of information via protoplasmic flux. As a result, sometimes \emph{Physarum} apparently chaotically abandons one direction to return to its previous path and explore that path some more. 

\begin{figure}[!tbp]
\centering
\includegraphics[scale=1]{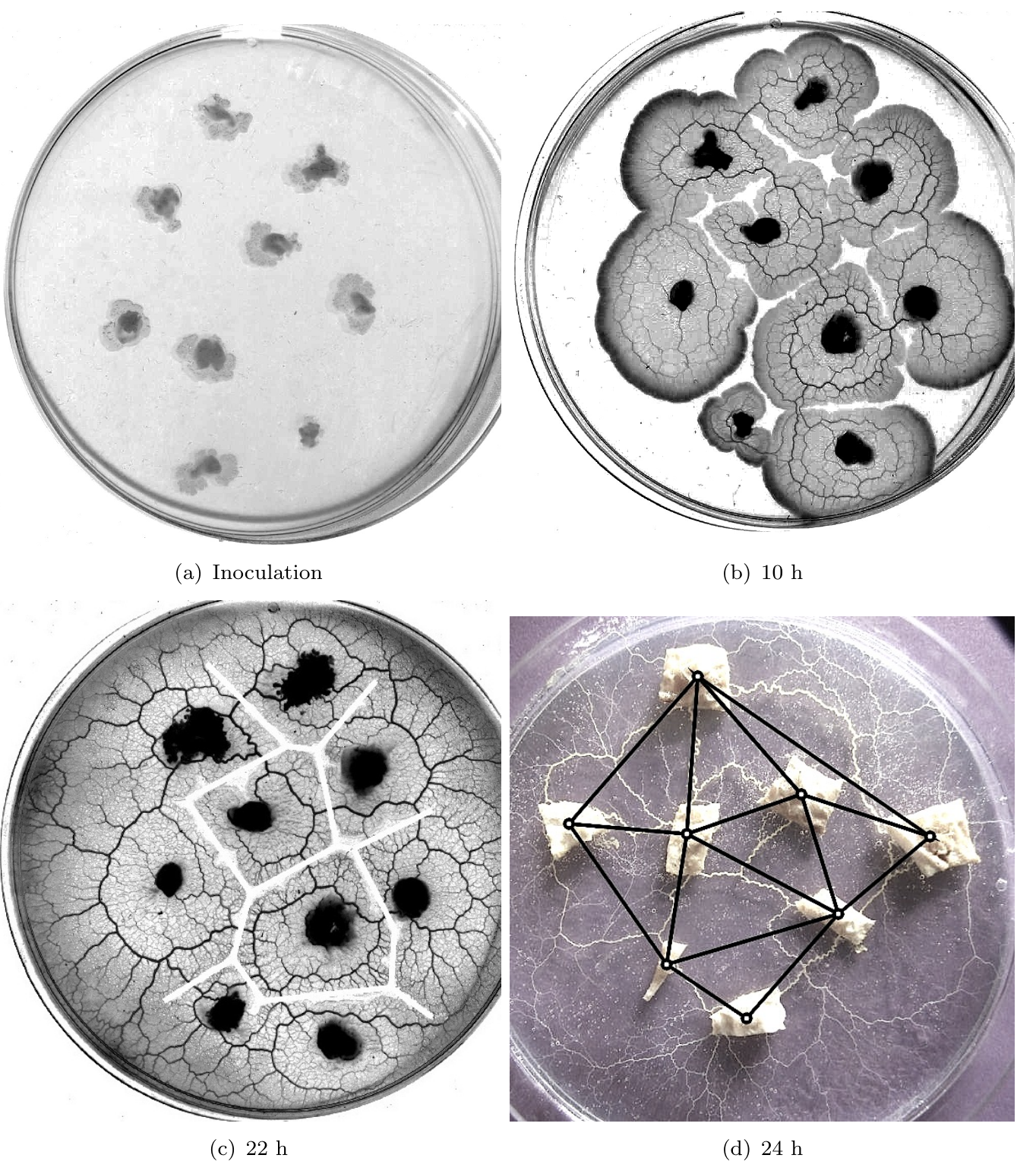}
\caption{Voronoi diagram and Delaunay triangulation developed during slime mould growth. 
(a--d)~Approximation of Voronoi diagram by slime mould  on nutrient agar gel. 
(a)~Sites of plasmodium inoculation represent planar data points to be sub-divided by edges of Voronoi diagram.
(bc)~Experimental snapshots of growing plasmodia.
(d)~Bisectors of  Voronoi diagram are represented by loci of substrate not occupied by plasmodium, bisectors computed by  classical technique are shown by straight lines.
(e)~Delaunay triangulation approximated by slime mould growing on non-nutrient substrate. Edges of \emph{Physarum}-computed triangulation are represented by protoplasmic tubes; edges computed by classical algorithm are solid lines.
Modified from~\cite{adamatzky_physarummachines}.
}
\label{pvoronoi}
\end{figure}

 Let us consider the effects of switching associations between concepts by considering two related mathematical structures: Voronoi diagram and Delaunay triangulation. 
 A planar Voronoi diagram of the set $\bf P$ is a partition of the plane into such regions that, for any element of $\bf P$, a region corresponding to a unique point $p \in {\bf P}$ contains all those points of the plane which are closer to $p$ than to any other node of $\bf P$.  Delaunay triangulation is a dual graph of Voronoi 
 diagram~\cite{shamos_preparata}.

Delaunay triangulation is a proximity graph, where any two planar nodes are connected by an edge if their local neighbourhood satisfy certain conditions. Voronoi diagram is not a proximity graph yet it is dual of Delaunay triangulation.  Two concept-structures are far from each other and not even humans (unless trained and aware) can spot the
relation between the diagram and triangulation. Slime mould \emph{P. polycephalum} finds the connection between 
the concepts. Plasmodium growing on a nutrient substrate from a single site of inoculation expands circularly as a typical diffusive or excitation wave. When two plasmodium waves encounter each other, they stop propagating. To approximate a Voronoi diagram with \emph{Physarum}, we physically map a configuration of planar data points by inoculating plasmodia on a substrate (Fig.~\ref{pvoronoi}a). Plasmodium waves propagate circularly from each data point (Fig.~\ref{pvoronoi}bc) and stop when they collide with each other (Fig.~\ref{pvoronoi}d). Thus, the plasmodium waves approximate a Voronoi diagram, whose edges are the substrate's loci not occupied by plasmodia (Figs.~\ref{pvoronoi}d).  

When \emph{Physarum} is given the same set of data (planar points represented by oat flakes colonised by plasmodium) but placed on a non-nutrient substrate. Being driven by chemo-attractants the plasmodium in each planar point develops just few localised active zones, which grow towards geographically neighbouring oat flakes (planar points). Thus the flakes become connected by enhanced protoplasmic tubes, which --- up to some degree of accuracy --- represent edges of the Delaunay triangulation (Figs.~\ref{pvoronoi}e)~\cite{adamatzky_ppl_2008}.

\section{Switching between conventional and unconventional ideations}

\begin{figure}[!tbp]
\centering
\includegraphics[scale=1]{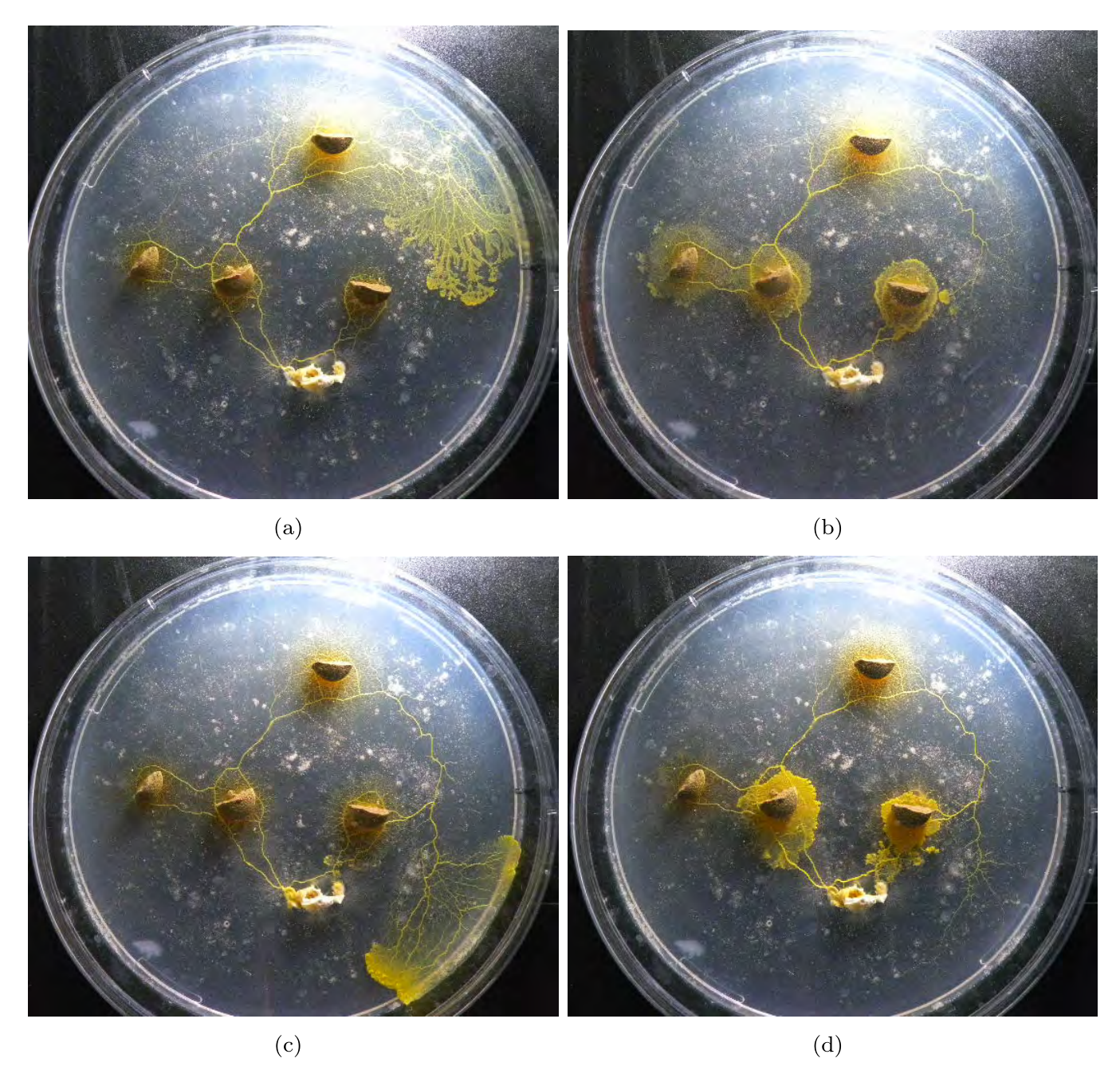}
\caption{Continued propagation of \emph{Physarum} after spanning of Nytol pills is constructed. 
See beginning of the dynamics in Fig.~\ref{tree}.}
\label{treecontinuation}
\end{figure}

Apparently spontaneous switching of growth modes was observed in \emph{Physarum} by 
Yamachiyo and colleagues~\cite{yamachiyo_2011}.  On few occasions we witnessed \emph{Physarum} deviating its 
growth away from source nutrients, which may be due to colonisation of nutrients by some other moulds or
non-comfortable level of humidity and illumination. Starting to sporulate when it is not illuminated and not `hungry' or 
trying to go in to its sclerotium phase when not drying would  be a very unconventional thing for \emph{Physarum} to do; however we never observed such rather radical forms of unconventionality.  

Continuing to solve a task when it is already solved could be classified as a way of flexible thinking. In  Fig.~\ref{tree} we  shown that give a planar set of sources of attractants \emph{Physarum} spans the set by a spanning tree. When all sources are spanned by a minimum-length acyclic graph the problem seems to be solved. Not for the slime mould, see 
Fig.~\ref{treecontinuation}. On reaching the northmost pill of Nytol \emph{Physarum} does not stop but propagates south 
(Fig.~\ref{treecontinuation}a) till re-colonises eastmost pill (Fig.~\ref{treecontinuation}b). It then continues moving along already established path forming a circular pattern of activity between initial inoculation site and Nytol pill  (Fig.~\ref{treecontinuation}cd).

\section{Generation of novel practically useful ideas}

`Novelty' for \emph{Physarum} relates to newly colonised territories. Thus \emph{Physarum} always aims to be novel. Practical usefulness is expressed in the slime mould's ability to solve complex (even for humans) problems with minimal resources, not necessarily in minimal time, and its ability to physically `think' around obstacles. 

Let us consider an example of a maze problem~\cite{adamatzky_2011}. A typical strategy for a maze-solving with a single device is to explore  all possible passages, while marking visited parts, till the exit or a central chamber is found.  
A central chamber of a maze is marked with chemo-attractant, e.g. emitted by sources of nutrients placed in the chamber. Plasmodium is placed in an outer chamber. 

\begin{figure}[!tbp]
\centering
\includegraphics[scale=1]{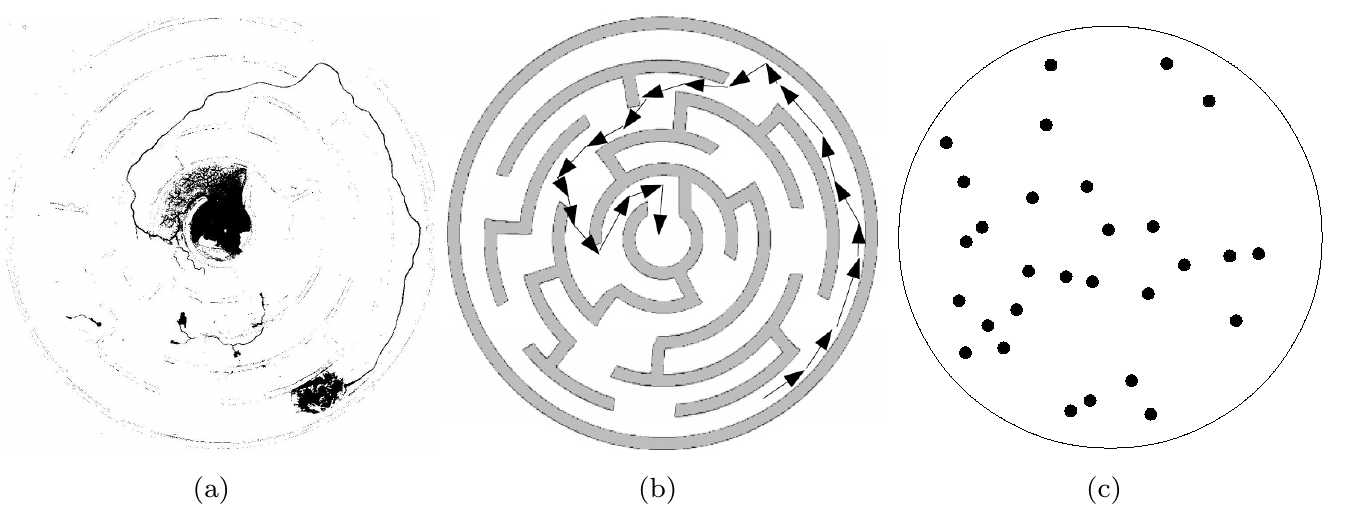}
\caption{Illustration of \emph{Physarum} creativity via maze problem~\cite{adamatzky_2011}. 
(a)~Photo of an experimental maze solution.
(b)~Scheme of plasmodium trajectory. 
(c)~Distribution of active zones.}
\label{maze}
\end{figure}
 
 After its inoculation the plasmodium started exploring its vicinity and at first generated two active zones propagating clock- and contra-clockwise (Fig.~\ref{maze}ab). Several active zones are developed to explore the maze (Fig.~\ref{maze}c). By the time diffusing chemo-attractants reached distant channels, one of the active zone already became dominant and suppressed another active zones. In example  shown active zone travelling contra-clockwise inhibited active zones propagating clockwise. The dominating active zone then  followed the gradient of chemo-attractants inside the maze, navigated along intersections of the maze's channels and solved the maze by entering its central chamber.  
 Distribution of active zones in Fig.~\ref{maze}c reflects patterns of \emph{Physarum} `thinking' while it solves the maze problem.

\section{Tendencies to functional non-conformity}

Functional non-conformity of \emph{Physarum} is manifested in its self-avoidance. As we discussed in example of Voronoi diagram in Sect.~\ref{remoteassociations} and \cite{adamatzky_physarummachines,adamatzky_parco} on a 
nutrient rich substrate plasmodial waves propagating from sites of inoculation do not interact. When two waves 
approach each other they stop. Loci of substrate at the boundary between growing patterns may remain unoccupied 
for a substantial but then some sporadical localised merging occurs wave-patterns originated from different sources of inoculation become connected. Having said that when two growing protoplasmic tubes collide they do quite often merge with each other. Tubes usually develop on a non-nutrient substrate. Thus we could speculate that non-conformity takes place in situations when nutrients are supplied in abundance. Deficiency of nutrients leads to conformism.

When nutrients in a substrate are exhausted or concentration of excreted metabolites becomes excessive \emph{Physarum} 
abandons the exhausted/contaminated region. The slime mould's protoplasmic tubes are sticking to substrate. Thus \emph{Physarum} could not just crawl away or retract his protoplasmic tubes. However, \emph{Physarum} can relocate its protoplasm by 
pumping it by peristaltic motion to parts of its body resting in more favourable regions. When protoplasm is pumped away from a tube the tube collapses yet remains in place and its external walls are clearly visible. 
Reid and colleagues (2012)
shown that \emph{Physarum} usually avoids its own abandon tubes: being place in a region with abandoned tube the slime mould propagates into not yet occupied domains.  The finding is trivial yet a useful piece of evidence support \emph{Physarum's} self-avoiding behaviour. 

The self-avoidance of \emph{Physarum} is quite remarkable because it seems like most creature do rather prefer follow
footpaths developed by others. Ants are the most known example. Most ants (not all) do develop their foraging by laying a trail of pheromones~\cite{camazine_2003}. The larger trails attract more ants and thus increase concentration of pheromones while smaller trail gradually evaporate. Both ants and \emph{Physarum} thus exhibit collective memory (mediated by the environment) but in the case of \emph{Physarum} this memory is not auto-catalytic (which, in ants is related to path length optimisation). Instead, the self-avoiding behaviour may relate to a distributed memory of previous occupancy (during nutrient foraging) and thus an implicit representation of the resources available in different regions of its habitat. The implicit knowledge of previously explored regions (coupled with diminishing sensory cues from depleted nutrients in these regions) ensures novelty in the selection of foraging direction and affords the plasmodium a mechanism by which it can sense when a region is no longer a good supply of nutrients, thus preventing cyclic and non-fruitful exploration of depleted regions.

 \section{Schyzotipy versus cognitive control}
 
 \begin{figure}[!tbp]
\centering
\includegraphics[scale=1]{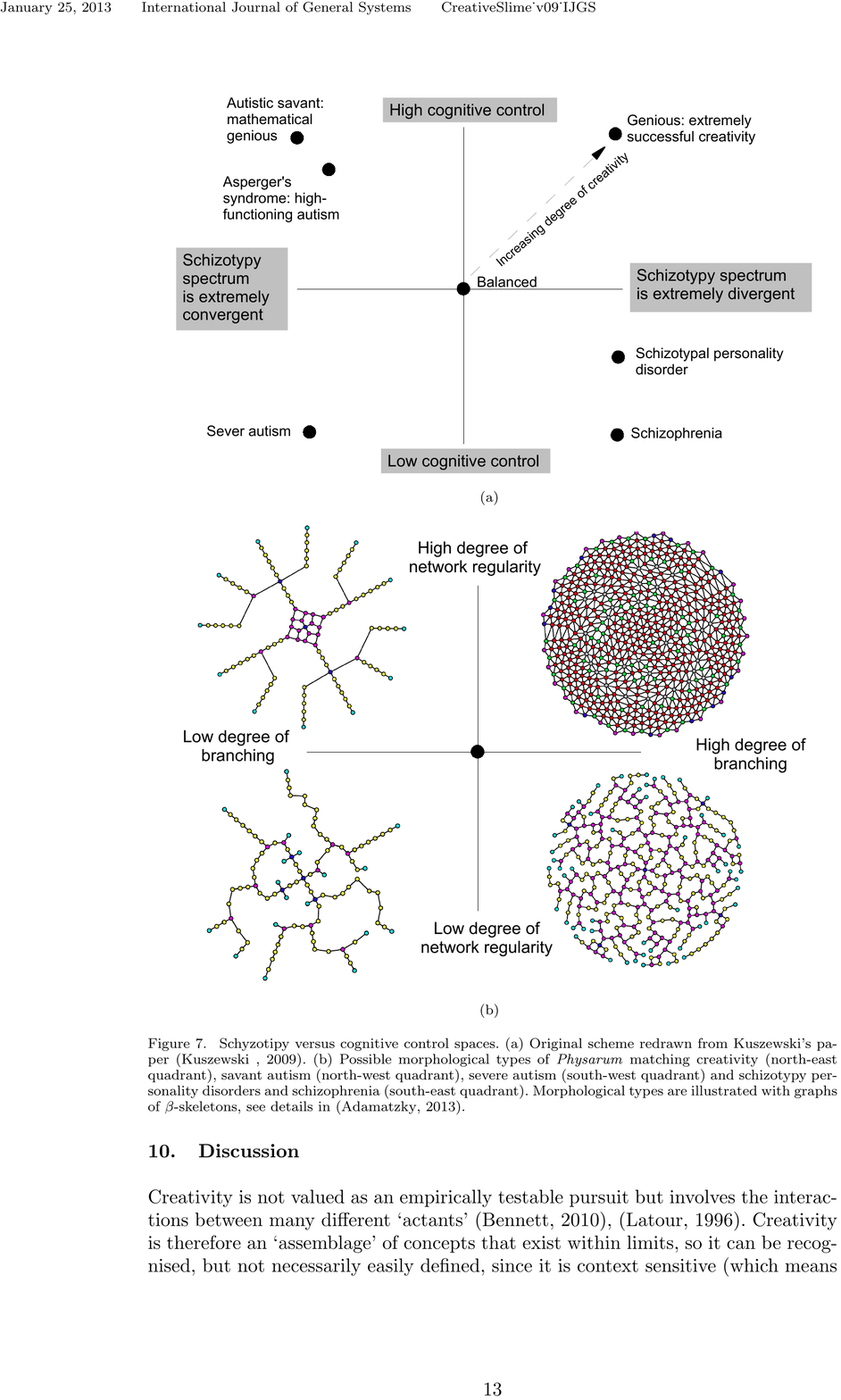}
\caption{Schyzotipy versus cognitive control spaces. (a)~Original scheme redrawn from Kuszewski's paper~\cite{kuszewski_2009}. (b)~Possible morphological types of \emph{Physarum} matching creativity (north-east quadrant), savant autism (north-west quadrant), severe autism (south-west quadrant) and schizotypy personality disorders and schizophrenia (south-east quadrant). Morphological types are illustrated with graphs of $\beta$-skeletons, see details in~\cite{adamatzky_growingskeletons}.
 }
\label{scheme}
\end{figure}


Cognitive control of divergent thinking is a requisite creativity. A person with extremely divergent thinking who is
unable to control these associations would be potentially classified as mentally ill. However, those who can fit their high schizotypy\footnote{A schizotypy is a range of personality characteristics ranging from normal to schizophrenia.} traits into rigorous cognitive frameworks may be classified as gifted or even genius.  Thus creativity could be positioned together with autism and schizophrenia in the same phase space 
(Fig~\ref{scheme}a). We can speculate that a degree of cognitive control is represented by a degree of regularity of slime mould's network. A degree of branching, or just an average number of neighbours, of a \emph{Physarum} network may be considered as a representation of a degree of schizotypy (Fig~\ref{scheme}b). Then `mathematical savant' slime mould grows a low branching highly symmetrical protoplasmic networks (north-west quadrant in Fig~\ref{scheme}b) and severely `autistic' slime mould develops highly asymmetric low branching networks  (south-west quadrant in Fig~\ref{scheme}b). `Schizophrenic' plasmodium of \emph{P. polycephalum} is a highly --- for planar proximity --- graph connected 
disordered network. High connected and highly ordered network, almost close to hexagonal packing of nodes, is a morphological structure of `creative' plasmodium.

 \section{Discussion}

Creativity is not valued as an empirically testable pursuit but involves the interactions between many different `actants' \cite{bennett_2010, latour_1996}. Creativity is therefore an `assemblage' of concepts that exist within limits, so it can be recognised, but not necessarily easily defined, since it is context sensitive (which means that different disciplines will value different aspects of creativity).

Slime mould of \emph{P. polycephalum} exhibits an `embodied thinking' emerging from its innate material interactions and mediated by the stimuli presented by its environment.  It `thinks' through tactility and it in a constant process of decision making. In this way the slime mould is similar to dancers and athletes who experience a kind of embodied thinking as they are habitually trained to use their bodies to express their feelings. One can speculate that with the increase of cognitive abilities afforded by higher-order living creatures' nervous systems, these creatures substitute some of the spatial and temporal aspects of their decision making (as employed by \emph{Physarum}) to more abstract representations and cognition. 


 \begin{figure}[!tbp]
\centering
\includegraphics[scale=1]{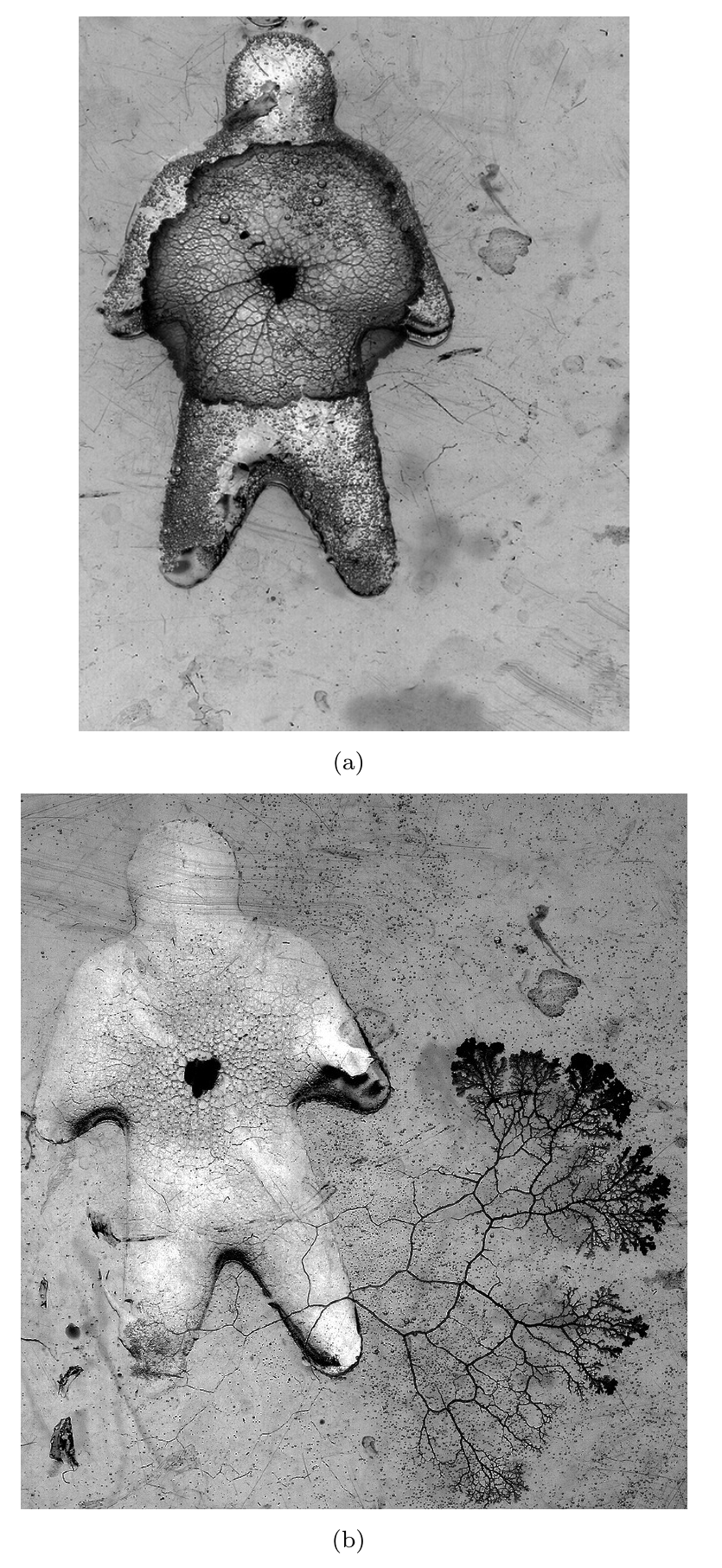}
\caption{Determination of the slime mould's morphology by substrate. Non-nutrient agar gel plate has a 
man-shaped insert filled with nutrient agar gel.  (a)~10~h after inoculation. (b)~22~h after inoculation.}
\label{substrate}
\end{figure}


A creativity needs context as much as an organism needs an environment. The creativity is shaped by different kinds of selection pressures. One person's `creativity' is another one's `mess'. Slime mould \emph{P. polycephalum} demonstrates this to a full extent through its relationship to `morphological meaning' in its environment, which is effected through a chemical `language'. \emph{Physarum's} behaviour can be interpreted as creative when the slime mould grows on a non-nutrient substrate with discrete configuration of attract sources distributed in the slime's environment yet it behaves in a dull, predictable, manner when colonises a homogeneous nutrient-rich substrate. This is experimentally evidenced in 
Fig.~\ref{substrate}.  We took a non-nutrient agar plate and cut off a man shape. We filled the man-shaped void nutrient agar gel 
and inoculated the slime mould inside the shape. \emph{Physarum} starts to grow as an omnidirectional pattern, circular wave,
inside the shape (Fig.~\ref{scheme}a). When the slime mould enters non-nutrient part, it changes its shape to a tree-like structure (Fig.~\ref{scheme}b).

We were using a formal set of criteria (divergent thinking, remote associations, flexibility in thinking, functional non-conformity)  for interrogation of the creative potential of slime mould. However, the value of the exercise 
is in the interrogation of the possibility, more than resolving the essence of the idea  which would, in many ways, be an impossible task. Mark Eli Kalderon (2005) observed the formal interest of philosophy in ÔfictionalismÕ, the view that a serious intellectual inquiry need not aim at truth.
This follows the publication of Hartry FieldÕs provocation that mathematics does not need to be true to be good~\cite{field_1980} and Bas van Fraassen's (1980)
sassertion that science does not seek truth, but empirical adequacy. Both authors proposed belief in the content of mathematics and science were not necessary for them to gain acceptance (full or tentative). Kalderon notes that the value of fictionalism is in the process of inquiry that it initiates. This precipitates a journey of intellectual discovery, which ends when the fiction reaches formal acceptance. 

We propose that the building blocks of creativity in a dynamic system without a formal nervous system can be considered via external relationships with 'morphological meaning'. As per  Csikszentmihalyi's criteria,  creativity is expressed where relationships between domain, field and person form an assemblage of interactions. This model not only provides a means of thinking about the generation of novelty in material systems but also deals with creativity without introducing new concepts and levels of complexity to define human-induced, subject/object relations. We suggest that the  further study of Physarum as a means of investigating 'spatialized' thinking may be a useful comparator in other 'cognitive' capacities such as, the formation of memory.

\section{Acknowledgement}

AA and JJ acknowledge support by  the  EU  research  project  ÒPhysarum Chip:   Growing Computers from Slime MouldÓ (FP7  ICT  Ref 316366).

\end{document}